\documentclass[conference]{IEEEtran}

\usepackage{amsmath}
\usepackage{graphicx}
\usepackage{subfig}
\usepackage{algorithm}
\usepackage{algorithmic}
\usepackage{color, soul}
\usepackage{cite}
\usepackage{url}
\sethlcolor{green}

\begin{document}
\title{In-Network Caching vs. Redundancy Elimination}

\author{\IEEEauthorblockN{Liang Wang\IEEEauthorrefmark{2}, 
Walter Wong\IEEEauthorrefmark{1},
Jussi Kangasharju\IEEEauthorrefmark{2}\IEEEauthorrefmark{3}}\\

\IEEEauthorblockA{\IEEEauthorrefmark{2}Department of Computer Science, University of Helsinki, Finland}
\IEEEauthorblockA{\IEEEauthorrefmark{1}School of Electrical and Computer Engineering, University of Campinas, Brazil}
\IEEEauthorblockA{\IEEEauthorrefmark{3}Helsinki Institute for Information Technology, University of Helsinki, Finland}
}

\maketitle

\begin{abstract}
  Network-level Redundancy Elimination (RE) techniques have been
  proposed to reduce the amount of traffic in the Internet.  and the
  costs of the WAN access in the Internet.  RE middleboxes are usually
  placed in the network access gateways and strip off the repeated
  data from the packets. More recently, generic network-level caching
  architectures have been proposed as alternative to reduce the
  redundant data traffic in the network, presenting benefits and
  drawbacks compared to RE. In this paper, we compare a generic
  in-network caching architecture against state-of-the-art redundancy
  elimination (RE) solutions on real network topologies, presenting
  the advantages of each technique. Our results show that in-network
  caching architectures outperform state-of-the-art RE solutions
  across a wide range of traffic characteristics and parameters.
  
\end{abstract}


\section{Introduction}
\label{s:introduction}

The fast growth of Internet bandwidth usage, mainly due to the
exponential increase in Internet videos (YouTube) and IPTV, has put
the Internet infrastructure under high pressure. According to a Cisco
survey~\cite{cisco:survey}, by 2014 the network traffic is expected to
approach 64 Exabytes per month, with videos accounting for more than
91\% of global traffic. Redundancy elimination (RE) techniques have
been proposed to handle the huge amount of data in the access
networks. Their main aim is to remove requests and/or responses of
redundant data in the network, reducing the traffic and costs in the
access network.

RE techniques can be classified into two kinds: (a) caching to remove
transfers, and (b) data replacement with a shim header. Former relies
on caching network-level objects and storing them temporarily in the
network. Caching techniques rely on redundancy of the
traffic~\cite{ager:cache,anand:sigmetrics}, implying that a large
portion of the network traffic is duplicated and could be cached for
later requests. Another incentive is that storage prices have
decreased faster than bandwidth costs \cite{ietf:decade}.


The second approach replaces redundant data with a shim header in an
upstream middlebox (usually close to the server) and reconstructing it
in a downstream middlebox before delivering it to the
client. Commercial products provide WAN optimization mechanisms
through RE in enterprise networks \cite{riverbed:wan, juniper:wan,
  bluecoat:wan}. Recently, RE has received considerable attention from
the research
community~\cite{anand:sigmetrics,anand:smartre,anand:cache,zohar:pack}. In~\cite{anand:cache},
the authors propose a network-wide approach for redundancy elimination
through deployment of routers that are able to remove redundant data
in ingress routers and reconstruct it in egress routers. However, they
also require tight synchronization between ingress and egress routers
in order to correctly reconstruct the packet and they also require a
centralized entity to compute the redundancy
profiles. In~\cite{zohar:pack}, the authors propose to use caches in
the local host and use prediction mechanisms to inform servers that
they have already the following redundant data. However, they are not
able to share the cached data among other nodes due to the local
characteristic of the cache.


Although both caching and RE have been around in the research
community, there has not been any thorough comparison in the
effectiveness of the two above-mentioned strategies: in-network
caching
vs. redundancy elimination. Work
in~\cite{Perino:2012:IRE:2342488.2342508} combines in-network caching
and RE, but limiting the applicability of the solution to a single
content source only.

In this paper, we perform a comparison between an in-network caching
architecture (INCA) and state-of-the-art RE solutions. Although INCA
models a generic network caching architecture, it is effectively
CCN-like~\cite{jacobson:ccn,wong:icc2011}. However, as we want to
understand the performance differences between caching and RE, we do
consider low-level protocol details.

We perform an extensive comparison, using real network topologies from
Rocketfuel, between INCA and RE. We have implemented the different
solutions on our testbed and compare them by running them on real
network topologies. We consider the position of a single ISP
interested in reducing its traffic both within and outside of its own
network.

Our key findings can be summarized as follows:
\begin{itemize}
\item In terms of reducing external network traffic, INCA is always
  superior when compared to ISP-internal RE
  solutions~\cite{anand:smartre}. End-to-end RE
  solutions~\cite{zohar:pack,anand:cache} can reduce external traffic,
  but are outside the control of the ISP; furthermore, they are not as
  effective as INCA.
\item In terms of reducing internal network traffic, INCA is in most
  cases clearly superior to state-of-the-art RE
  solutions\cite{anand:smartre}, with at least 50--65\% improvements
  in internal traffic reduction.
\end{itemize}



The organization of this paper is as follows. Section \ref{s:background} 
presents the background information and related work about in-network
caching and redundancy elimination solutions. Section
\ref{s:architecture} introduces the in-network caching
architecture (INCA), describing its main features. Section
\ref{s:evaluation} presents the evaluation methodology, and the
comparison results between INCA and RE solutions. Finally,
Section~\ref{s:conclusion} summarizes the paper.

\section{Background}
\label{s:background}


\subsection{Caching}
\label{sec:caching}


%
%

Recently, information-centric networking (ICN),
e.g.,\cite{jacobson:ccn,psirp,koponen:dona}
has emerged as a more general, network-wide caching solution. In ICN,
content caches in the network (e.g., in routers) store content that
passes through them and if they see requests for the same content, they
are able to serve it from their cache. 
INCA is essentially an ICN architecture, but our intention is not to
provide yet-another-ICN-architecture. Instead, INCA simply considers
the key features of ICN architectures, namely caching and routing
towards some point of origin for content, and ignores practical,
low-level protocol details. INCA draws inspiration from
CCN~\cite{jacobson:ccn} and our previous work~\cite{wong:icc2011}, but
does not specify low-level behavior.


Other caching proposals also exist. Cache-and-Forward (CNF)
\cite{paul:cache} is an in-network caching architecture where routers
have a large amount of storage. These routers perform content-aware
caching, routing and forwarding \textit{packet} requests based on
location-independent identifiers, similar to CCN.


\subsection{Redundancy Elimination}
\label{sec:redund-elim}

Modern RE schemes use a fingerprint-based data stripping model.  Nodes
generate a set of fingerprints for each packet in transit, where each
fingerprint can be generated over a pre-defined block size. Upon
detecting a cached fingerprint, the upstream node replaces the data by
a fingerprint and the downstream node replaces the fingerprint with
the original data, reducing the overall data transmission over the
network. As described in~\cite{spring:re}, both upstream and
downstream nodes need to be strongly synchronized in order to work
correctly. A similar approach is presented
in~\cite{aggarwal2010endre}.

Work in~\cite{anand:cache, anand:smartre} proposes to extend the RE
technique to the whole network, i.e., to make RE as a basic primitive
for Internet.
The main idea is to collect redundancy profiles
from the network and use a centralized entity to compute paths between
destinations within an ISP with higher RE capabilities. Therefore,
data going through these networks have higher RE footprint reduction
than going to other paths in the network. Despite the improved RE
capacity, it still requires strong synchronization between the
upstream and downstream routers in order to work properly.

A third RE approach~\cite{zohar:pack} was recently proposed to
overcome the synchronization issue in order to be deployed in
data-center networks. As cloud elasticity favors the migration and
distribution of work among a set of nodes, it is hard to set up the
synchronization between two fixed nodes. Therefore, the main idea
of~\cite{zohar:pack} is to create a local cache together with a
predictive mechanism to acknowledge already cached data to the
server. In this scenario, the service sends a predictive
acknowledgement to the server informing that the requested data is
already present in the client, thus, removing the redundant
data. Despite the improvement over the fixed node requirement, the use
of local storage prevents the sharing among other nodes, increasing
the overall sharing capacity and hit ratio. Therefore, the RE is not
network wide, but for redundant data that may be requested again in
the local node.

\section{INCA: In-Network Caching Architecture}
\label{s:architecture}

INCA focuses on the following key aspects of ICN architectures:
routing requests for content towards a known point, caching of
content, and forwarding responses back to the requesting entity. 
This model is similar to CCN~\cite{jacobson:ccn}.

\subsection{Basic Model}
\label{sec:basic-storage-}

The basic in-network caching mechanism is performed by a
\textit{content router} (CR). A CR is a data forwarder similar to a
regular router, but has some internal memory that can be used to store
data in transit. Each piece of content has a \emph{chunk ID} as its
permanent identifier from a cryptographic hash function.  Any CR on
the path between a server and clients caches the data in its
memory. Further requests can be served by the local copy in the
CR.  For a further discussion on this model and its limitations, we
refer the reader to~\cite{wong:globecom2012}.




\subsection{Caching in CRs}
\label{sec:caching-crs}

As in~\cite{wong:globecom2012}, we use three admission policies for
deciding which content a CR caches. 

\begin{itemize}

\item \textbf{ALL} admits all objects into the storage at the CR. In
  other words, every object that transits through the CR is taken into
  storage and another object is possibly evicted. This is the typical
  behavior of web caches.

\item \textbf{Cachedbit}~\cite{wong:globecom2012} sets one bit in the
  CR header to indicate whether a given piece of content has already
  been cached or not, preventing duplicated content along the same
  path. If the path between the client and server is $n$ hops, then a
  CR will cache the content with probability $1/n$ and once the
  content is cached, downstream CRs will not cache it, with the
  exception of the last CR on the path which will always cache it (see
  Section~\ref{ss:experimental-results} for an explanation).

\item \textbf{Neighbor Search (NbSC)}~\cite{wong:globecom2012} works
  like \textit{Cachedbit}, but if a CR encounters a miss, it will
  query neighboring CRs for that piece of content. CRs periodically
  exchange Bloom filters of their contents with their neighbor
  CRs. Please see~\cite{wong:globecom2012} for details about the size
  of Bloom filters, exchange frequency, and query radius.

\end{itemize}

We use Least Recently Used policy to decide what to evict when the
storage at the CR is full.  The results in~\cite{wong:globecom2012}
showed that a Cachedbit-like admission policy is needed to get good
caching performance, but that the addition of NbSC gives a
considerable boost in reducing network traffic.

\section{Evaluation \& Experimental Results}
\label{s:evaluation}

We chose 4 real-world networks from
Rocketfuel~\cite{SpringN:Rocketfuel}: Exodus, Sprint, AT\&T and NTT,
and performed a set of experiments using different cooperative caching
strategies. Table~\ref{tab:topologies} shows an overview of the
networks. All the experiments are performed on our department cluster
consisting of Dell PowerEdge M610 nodes. Each node is equipped with 2
quad-core CPUs, 32GB memory, and connected to 10-Gbit network. All the
nodes run Ubuntu SMP with 2.6.32 kernel.


\begin{table}[!tb]
  \centering
  \begin{tabular}{|c|c|c|c|}
    \hline
    Network & Routers & Links & \# of POPs \\
    \hline
    Exodus & 338 & 800 & 23 \\
    \hline
    Sprint & 547 & 1600 & 43 \\
    \hline
    AT\&T & 733 & 2300 & 108 \\
    \hline
    NTT & 1018 & 2300 & 121 \\
    \hline
  \end{tabular}
  \caption{Topologies used in experiments}
  \label{tab:topologies}
  \vskip -5mm
\end{table}

Our focus in comparison was to compare ICN-like in-network caching
represented by our INCA architecture with state-of-the-art RE
solutions. As points of comparison from the RE space, we selected
three solutions. We picked SmartRE~\cite{anand:smartre} because it
represents a solution internal to a single ISP, much like INCA could
be deployed. As examples of end-to-end RE, we selected
EndRE~\cite{aggarwal2010endre} and PACK~\cite{zohar:pack}.

We implemented SmartRE on top of our software router architecture and
used an LP solver to follow the behavior defined
in~\cite{anand:smartre}.  For EndRE and PACK, we simply compare the
performance numbers from the original papers to the numbers of INCA
from our experiments.

\subsection{Methodology}
\label{sec:methodology}

We use $10^5$ distinct chunks  of 1~KB as
our data set in the experiments. The popularity follows Zipf
distribution and the popularity of the $i$th most popular chunk is
proportional to $1/i^{\alpha}$; we use 0.7, 0.9 and 1.1 as $\alpha$
value.


Content popularity on the Internet is known to follow a Zipf- or a
power law distribution~\cite{breslau:zipf,cha:zipf}. However, in
practice one piece of content would consist of several (tens,
hundreds, or thousands of) packets. In many cases, there are strong
dependencies between packets, i.e., packets belonging to the same
piece of content would often be requested together. We have decided to
use only single chunks to represent all packets of a piece of content
for the reason of reducing the number of parameters to be
explored. There is little available information about the distribution
of size of content pieces, making it difficult to plug in convincing
size distributions. Because of the strong dependencies, one chunk in
our experiment would translate to several packets in the real world,
and thus the cache sizes would need to be adapted to match that.

%

The experiment followed the style in~\cite{anand:smartre}. We placed
clients and servers at a POP, to represent all the potential servers
or clients behind that POP. CRs were installed on every router. We
chose top 20 POPs with highest degree as servers, and rest of the POPs
as clients. Exodus has only 10 servers due to its small size. A client
keeps requesting the chunks from different servers. We experimented
two traffic patterns: constant and a gravity model, similar
to~\cite{anand:smartre}, but differences were negligible.

The metrics we investigated were:
\begin{itemize}

\item \textbf{Hit rate:} What fraction of requests was served by
  CRs. Hit rate in our context measures the savings in external
  traffic from the providers.

\item \textbf{Content locality:} We analyzed the number of hops needed
  to get the content to evaluate how the different algorithms are able
  to get content close to the users. Average hop bears a relationship
  to the access latency.

\item \textbf{Footprint reduction:} Network footprint is the product
  of the amount of data and the network distance from which the data
  was retrieved. It measures the amount of internal traffic reduction,
  i.e., a smaller footprint (larger reduction) means less traffic
  within the ISP's network. This metric was used also
  in~\cite{anand:smartre} and forms the basis of comparison between
  INCA and SmartRE.

\end{itemize}

Our goal was to see what, if any, performance differences there are
between INCA and the RE solutions, and determine the causes of the
differences. In order to better understand INCA's behavior, we
evaluated it very closely in terms of where it places the content and
how it uses it.

Note that hit rate does not apply to any of the RE solutions we
used. SmartRE reduces only internal traffic, but has no effect on
external traffic; in essence it has zero hit rate. End-to-end RE
solutions reduce traffic across the whole network, but an intermediate
ISP has no control over it, thus there is no ``hit rate'' since the
ISP must transit all traffic that it sees, although the amount of
traffic is less than without an RE solution.  Likewise, content
locality in SmartRE is subsumed by footprint reduction and locality
does not apply to end-to-end RE since all content always comes from
the origin, although with eliminated redundancy.

We repeated experiments to eliminate variability in the results.
Confidence intervals were very tight even after 5 repetitions and for
clarity reasons we do not show them.

In the following, we first present INCA's performance in terms of hit
rate and content locality; as discussed above these do not apply to RE
solutions. Then we show the comparison between INCA and SmartRE for
footprint reduction and compare INCA against end-to-end RE solutions.

\subsection{Experimental Results}
\label{ss:experimental-results}

\subsubsection{Hit Rate}
\label{sss:hit-rate}

\begin{figure*}[!tb]
  \subfloat[Sprint]{\label{fig:hit-sprint}\includegraphics[width=6.2cm]{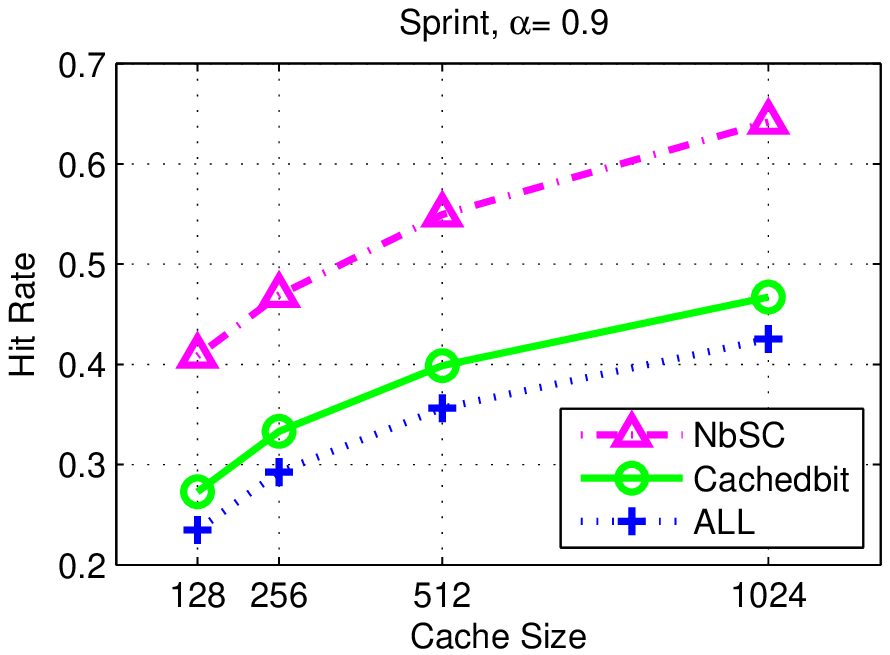}}
  \subfloat[AT\&T]{\label{fig:hit-att}\includegraphics[width=6.2cm]{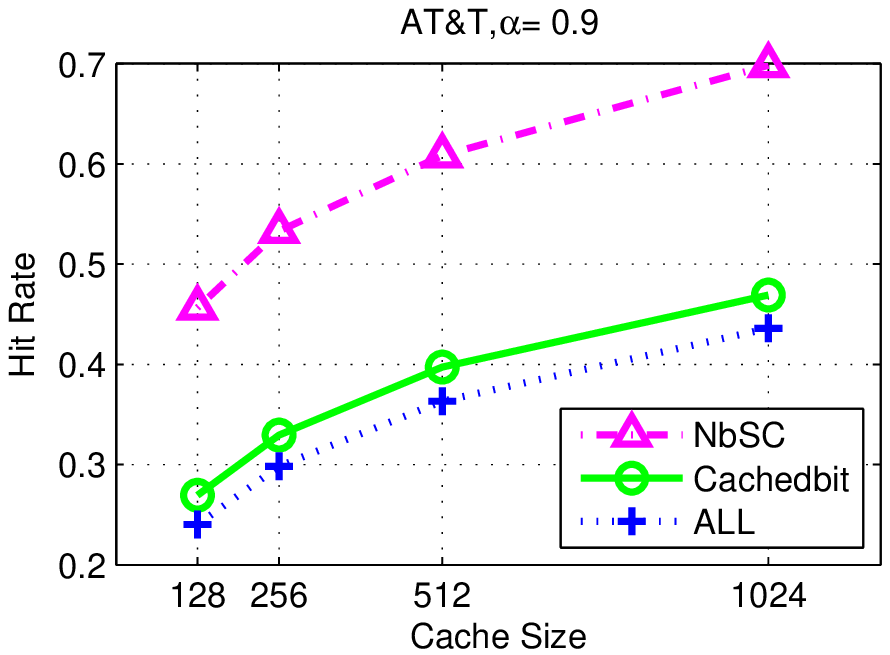}}
  \subfloat[NTT]{\label{fig:hit-ntt}\includegraphics[width=6.2cm]{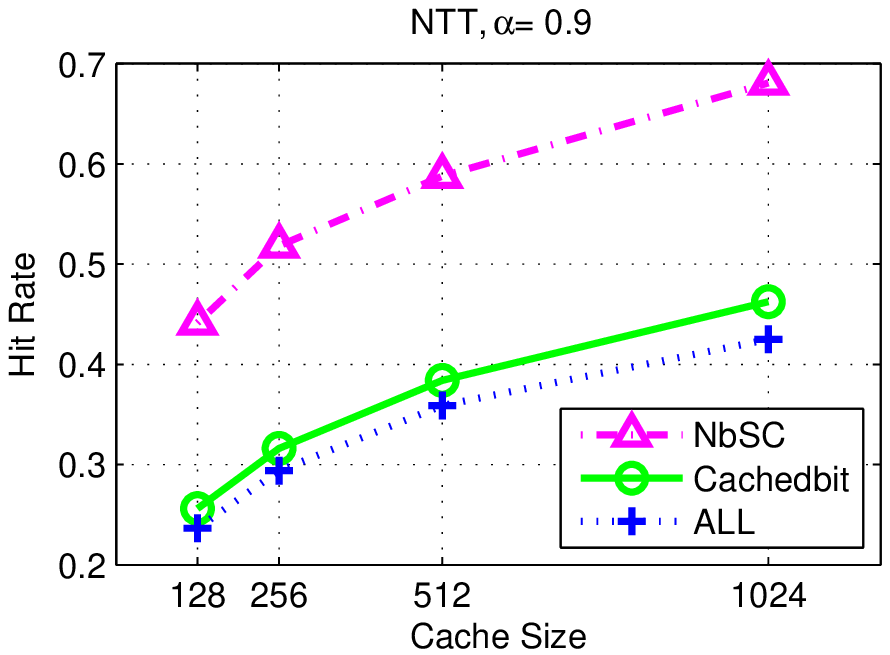}}
  \caption{Hit rate vs. network topology on POP level.}
  \label{fig:hitrate}
  \vskip -5mm
\end{figure*}

Figure~\ref{fig:hitrate} shows the hit rates in three networks we
studied. (Exodus yielded similar results and for reasons of space we
omit showing them.) On the x-axis we show the number of chunks each CR
could store and the y-axis shows the hit rate. Each graph shows 3
curves, one for each admission policy. Recall that we had $10^5$ chunks in the experiment, meaning that even with 1000 chunks of
storage per CR, one CR can store only about 1\% of the total amount of
chunks.


Neighbor Search has the highest hit rate and Cachedbit is better than
ALL policy. The results here are shown for $\alpha = 0.9$; for $\alpha
= 0.7$ or $1.1$ the ranking was the same, but the absolute values were
lower or higher, respectively.

Even though the networks vary considerably in size, it actually turns
out that the network paths between clients and servers are roughly
similar in length in all three networks. This means that in all
networks the caching capacity on a path is similar, hence getting
similar hit rates is to be expected. This is one of the key findings
in our work regarding caching performance: \emph{Caching performance
  of a CR network depends mainly on the path lengths and network
  topology rather than the absolute number of CRs in the
  network}. There is some additional evidence in previous studies on
Rocketfuel data~\cite{EumS:BioInspired} to suggest that the different
networks share some graph theoretical properties. Exactly which
properties are important for caching (besides path length) and how
they affect performance of caching networks is left for further study.



Both NbSC and Cachedbit show clear gains over the ALL policy. This
demonstrates the importance of not wasting storage space as is done by
the simple ALL policy, which always admits every chunk into a CR. In
contrast, the other two attempt to ensure that at most one copy of
a chunk is created (per path; recall that NbSC uses Cachedbit to
decide where to cache). Since both are probabilistic, it is possible,
that no copies are created. 

\subsubsection{Content Locality}
\label{sss:content-locality}

\begin{figure*}[!tb]
  \centering \subfloat[POP-level, 128
  chunks]{\includegraphics[width=6.2cm]{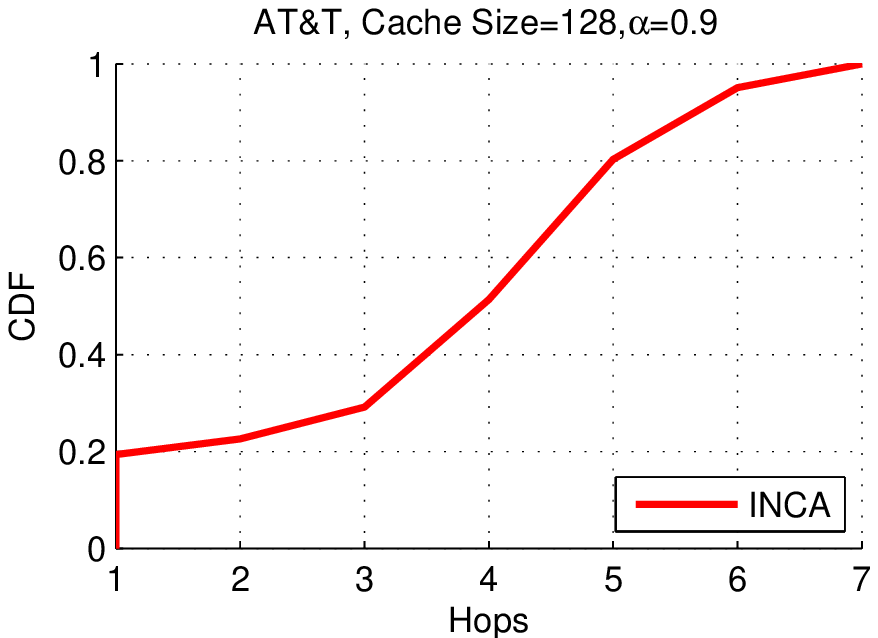}}
  \subfloat[POP-level, 512
  chunks]{\includegraphics[width=6.2cm]{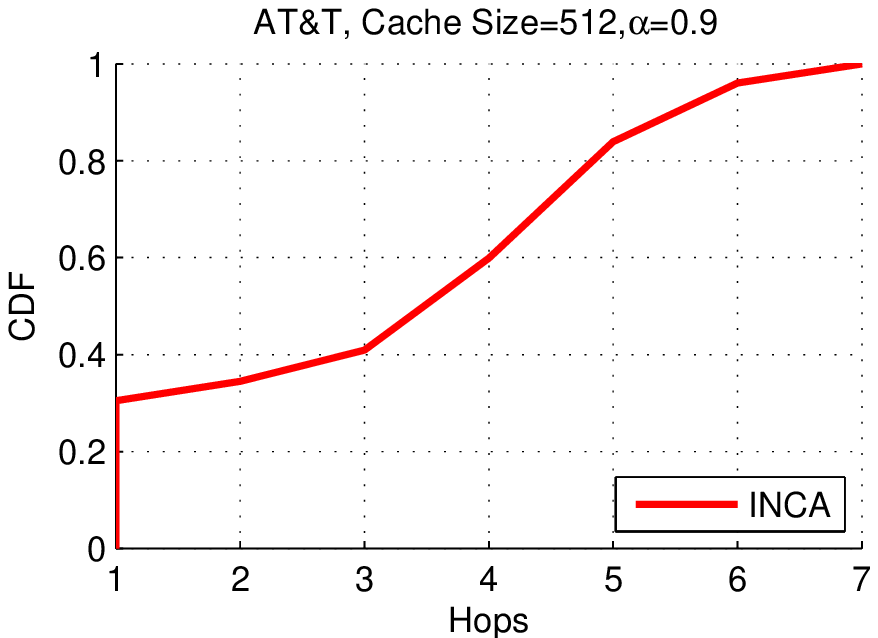}}
  \subfloat[Router-level, 128
  chunks]{\includegraphics[width=6.2cm]{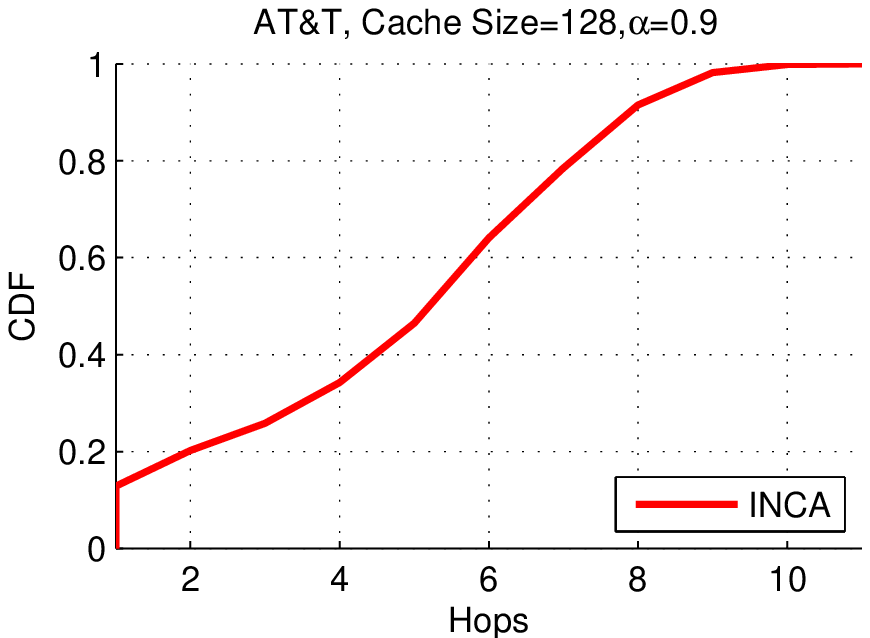}}
  \caption{CDF of number of hops to content in AT\&T network for
    $\alpha = 0.9$}
  \label{fig:hops0.8}
  \vskip -5mm
\end{figure*}

We also investigated how well the different algorithms were able to
get content close to the users. Figure~\ref{fig:hops0.8} shows the CDF
of the number of hops between CRs needed to get the content in the
AT\&T network with 128 chunks of storage per CR for POP-level and
router-level topologies and 512 chunks for POP-level topology. We only
plot the line for the Cachedbit strategy. NbSC uses the same placement
strategy and ALL was typically slightly worse than Cachedbit. Other
networks yielded similar results.





As expected, more storage per CR allows content to be cached closer to
the clients. On the router-level topology, the paths are slightly
longer, but the overall shape of the curve is similar to POP-level
curves.

The slow increase of the CDF indicates that about 30--40\% of the
cache hits happen in the first 3 POP-level hops. Partial explanation
of this is the experiment setup where we have clients behind every
non-server POP. This means that a POP that is the second hop for one
client is often also a first hop for some other client. In the end,
these clients end up fighting for cache space and typically the
clients closer to the POP end up getting their most popular content
there. Clients for whom that POP is the second or further hop, are
therefore less likely to find ``new'' content there, i.e., content
that was not already cached in the first POP. As the paths get longer,
this effect seems to wane and the CDF starts a faster increase.



We will discuss footprint reduction below and directly compare INCA
with SmartRE.

\subsection{INCA vs. SmartRE}
\label{sec:inca-vs.-smartre}

SmartRE~\cite{anand:smartre} uses two network elements, the redundancy
profiler and the redundancy-aware route computation element. The
redundancy profiler collects in-transit data statistics in order to
create a profile of the most popular data to be the ones to be cached
in the routers. The redundancy aware route computation computes the
paths based on the content stored in the network in order to optimize
the redundancy elimination of the network by solving a linear
programming (LP) problem. The benefit of such centralized element is
the fact that it knows the complete topology and makes it possible to
compute a good result for the RE. A totally decentralized SmartRE
model is not possible since there must be an entity controlling the
synchronization between these points.

SmartRE reduces the network footprint, because the caches between the
ingress and egress store parts of the data and the ingress simply
indicates which parts a cache is to substitute in a packet. There is
no effect on external traffic. The LP solver knows the redundancy
profile of the traffic and calculates a caching manifest which
indicates which parts of which packets should be decoded at which
caches. There is a very strong link between the total amount of
storage in the network and the length of the sampling period which
defines how long traffic is observed to compute the redundancy
profile.  According to~\cite{anand:smartre}, sampling periods on the
order of a few tens of seconds are to be expected to be reasonable.

We implemented SmartRE on top of our CR testbed. We noticed that
SmartRE, or rather the LP defined in~\cite{anand:smartre}, is very
sensitive to the parameters in the model. Small deviations often lead
to large differences in performance, typically for the worse. We were
able to determine parameters for what corresponds to the settings
in~\cite{anand:smartre} and calculated the footprint reductions for
the same traffic as with INCA. These ideal footprint reductions are
shown in Table~\ref{tab:smartre-footprint-ideal}.

\begin{table}[!tb]
  \centering
  \begin{tabular}{|l|c|}
    \hline
    \textbf{Network} & \textbf{FP Reduction} \\
    \hline
    Exodus & 27.55\% \\
    Sprint & 28.79\% \\
    AT\&T & 31.59\% \\
    NTT & 30.45\% \\
    \hline
  \end{tabular}
  \caption{SmartRE footprint reductions in different networks under
    ideal conditions}
  \label{tab:smartre-footprint-ideal}
  \vskip -5mm
\end{table}

\begin{figure*}[!tb]
  \subfloat[Sprint]{\label{fig:hit-sprint}\includegraphics[width=6.2cm]{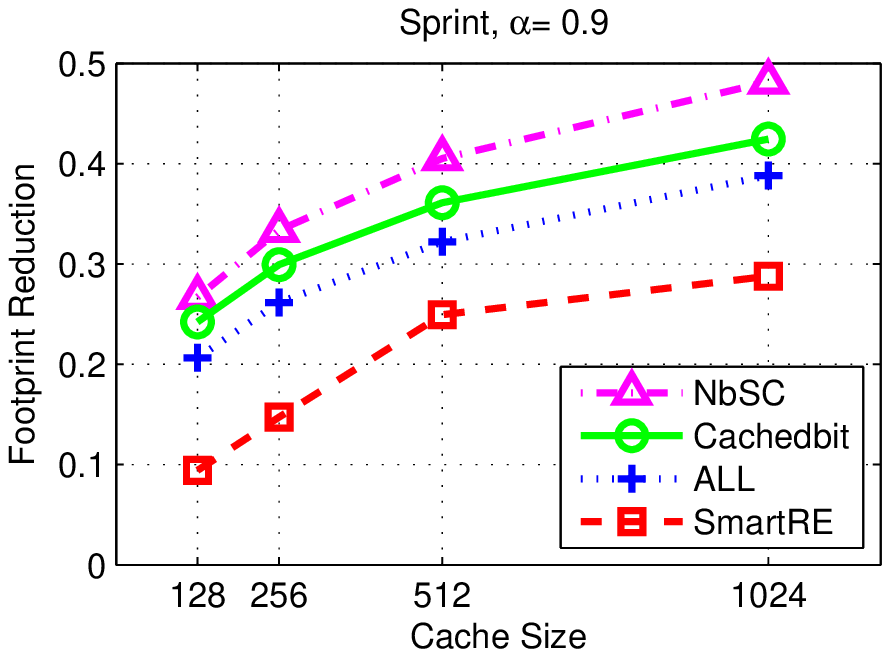}}
  \subfloat[AT\&T]{\label{fig:hit-att}\includegraphics[width=6.2cm]{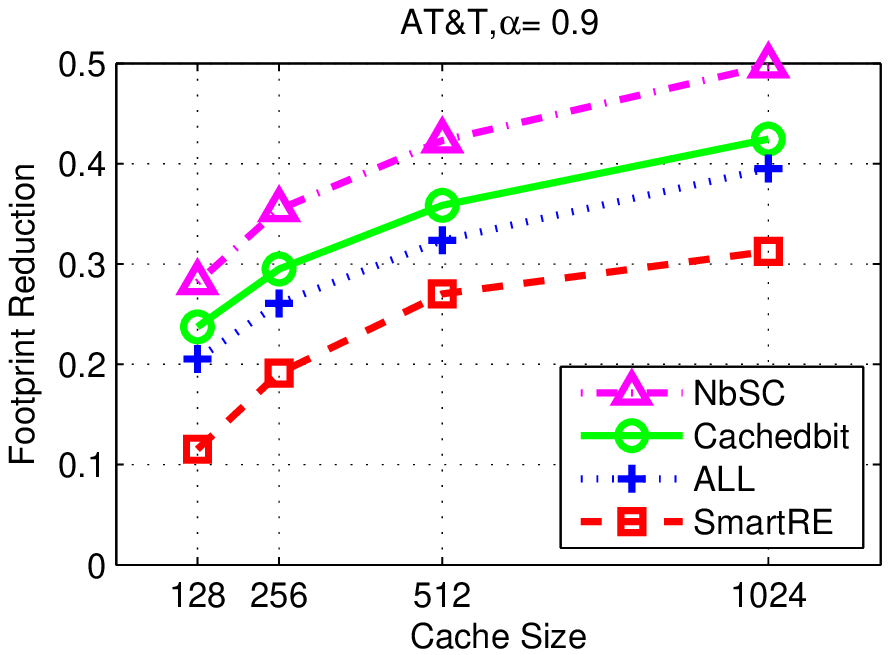}}
  \subfloat[NTT]{\label{fig:hit-ntt}\includegraphics[width=6.2cm]{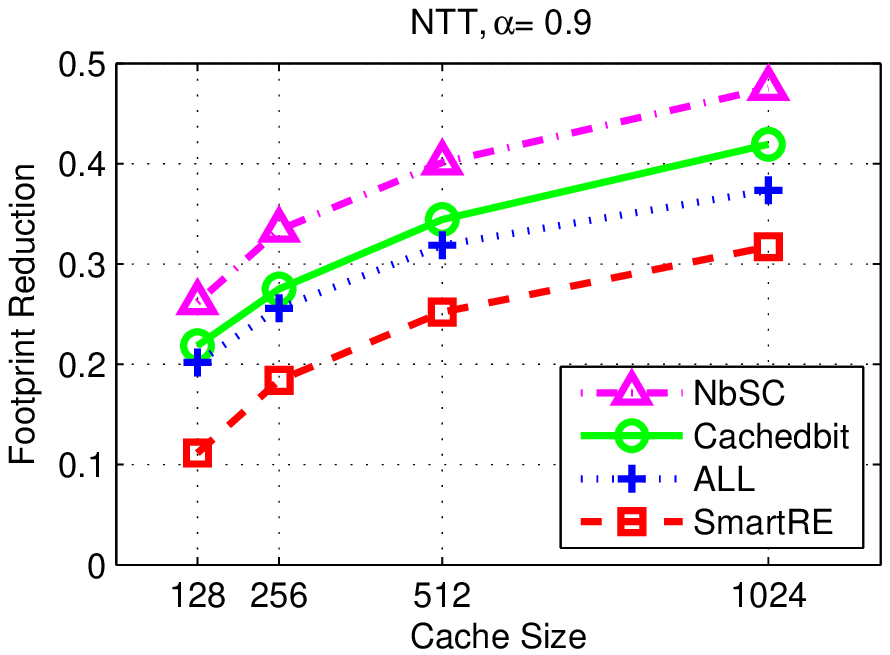}}
  \caption{Comparing footprints of INCA and SmartRE}
  \label{fig:footprint}
  \vskip -5mm
\end{figure*}

Figure~\ref{fig:footprint} shows the internal traffic reduction as
measured by the network footprint reduction. The y-axis shows the
fraction of internal traffic that was reduced by the caches in the
CRs. As with the other metrics, the differences between the three
admission policies are small. Again, NbSC is clearly superior to
Cachedbit which, in turn, is clearly superior to the ALL
policy. Footprint reduction is the reason why we tweaked Cachedbit to
create a copy of the chunk at the CR closest to the client. Without
the additional copy, ALL-policy is better at footprint reduction than
Cachedbit. We observed that this additional copying drops the hit rate
by a negligible amount, but raises the footprint reduction
considerably.

Contrasting the numbers in Table~\ref{tab:smartre-footprint-ideal}
with the INCA footprint reductions in Figure~\ref{fig:footprint}, we
see that they are similar in value. For small INCA cache sizes,
SmartRE yields a higher reduction, whereas for larger cache sizes,
INCA has the upper hand. However, even for very modest cache sizes,
NbSC is able to achieve an equal footprint reduction to SmartRE and
for large cache sizes, the footprint reduction is improved by
50--65\%. Cooperative caching is therefore much more efficient at
reducing internal traffic than SmartRE.

Recall that our INCA experiments considered one chunk to represent one
file, whereas in the SmartRE experiments, a chunk is one packet. This
means that the footprint reduction numbers cannot be directly compared
since traffic is different in the two cases. However, based on the
numbers presented in~\cite{anand:smartre}, we can infer a mapping
between SmartRE and INCA experiments. In~\cite{anand:smartre} it is
shown that SmartRE gets close to its ideal performance with 6~GB of
storage per router. Assuming the same 6~GB of storage per CR, the case
of 1024 chunks of storage, where 1 chunk equals 1 file, would imply
the average file size to be about 6~MB. If the content is a mixture of
text, images, and short videos, this seems like a reasonable, if not
even conservative, number. (For content consisting mainly of larger
videos, this would not be sufficient.)

We ran experiments with SmartRE where we took the ideal cache size
used to obtain the numbers for
Table~\ref{tab:smartre-footprint-ideal}, and set it to $1/2$, $1/4$,
and $1/8$ of that value. For each case, we then ran the experiment to
obtain the reduction in footprint. This allows us to plot the INCA and
SmartRE footprint reductions on the same x-axis, shown in
Figure~\ref{fig:footprint}. This confirms that INCA is more efficient
in reducing internal traffic in the network. The additional reduction
in traffic varies between almost 200\% for small caches and 50\% for
large caches.


Cachedbit is similar to the heuristic ``Heur1''
from~\cite{anand:smartre} in how it attempts to place the
content. In~\cite{anand:smartre}, the performance of these two
heuristics was found lacking when compared to the SmartRE algorithm
with its centralized controller deciding on what to cache where. If
the same translates to an INCA caching network, a centralized
controller deciding on placement of chunks in CRs would be a superior
choice. However, similar placement problems are often
NP-complete~\cite{QiuL:ReplicaPlacement}, although some
simplifications are likely to yield a linear program. We have not
considered a central placement agent in INCA, although it could be
included in future work.

An important difference is that INCA is able to share cache space
between clients, whereas SmartRE has fixed buckets for each
ingress-egress flow. This gives INCA more possibilities in exploiting
the cached data, thus reducing footprint and improving hit
rate. \emph{We believe this sharing of cache space between all client
  and server pairs is what gives INCA an advantage over SmartRE.}
Contrasting our results to the single server case presented
in~\cite{Perino:2012:IRE:2342488.2342508} is part of our future work.


Comparing INCA with SmartRE, we come to the following conclusions:

\begin{itemize}
\item For external traffic reduction, INCA is always superior, because
  SmartRE has no effect on external traffic.
\item For internal traffic reduction, performance of INCA (with
  neighbor search) is in most cases clearly superior, up to 50--65\%
  more reduction in internal traffic. However, the differences depend
  on how the mapping between cache sizes is done and the file size
  distribution, thus in different environments the results could be
  different. 
\end{itemize}

However, in our experimental environment INCA with neighbor search is
far more effective in reducing both internal and external traffic.

\subsection{INCA vs. End-to-End RE}
\label{sec:inca-vs.-end}

\begin{figure}[!tb]
  \centering
  \subfloat[128 chunks of storage]{\label{fig:inca-vs-endre-128}\includegraphics[width=7cm]{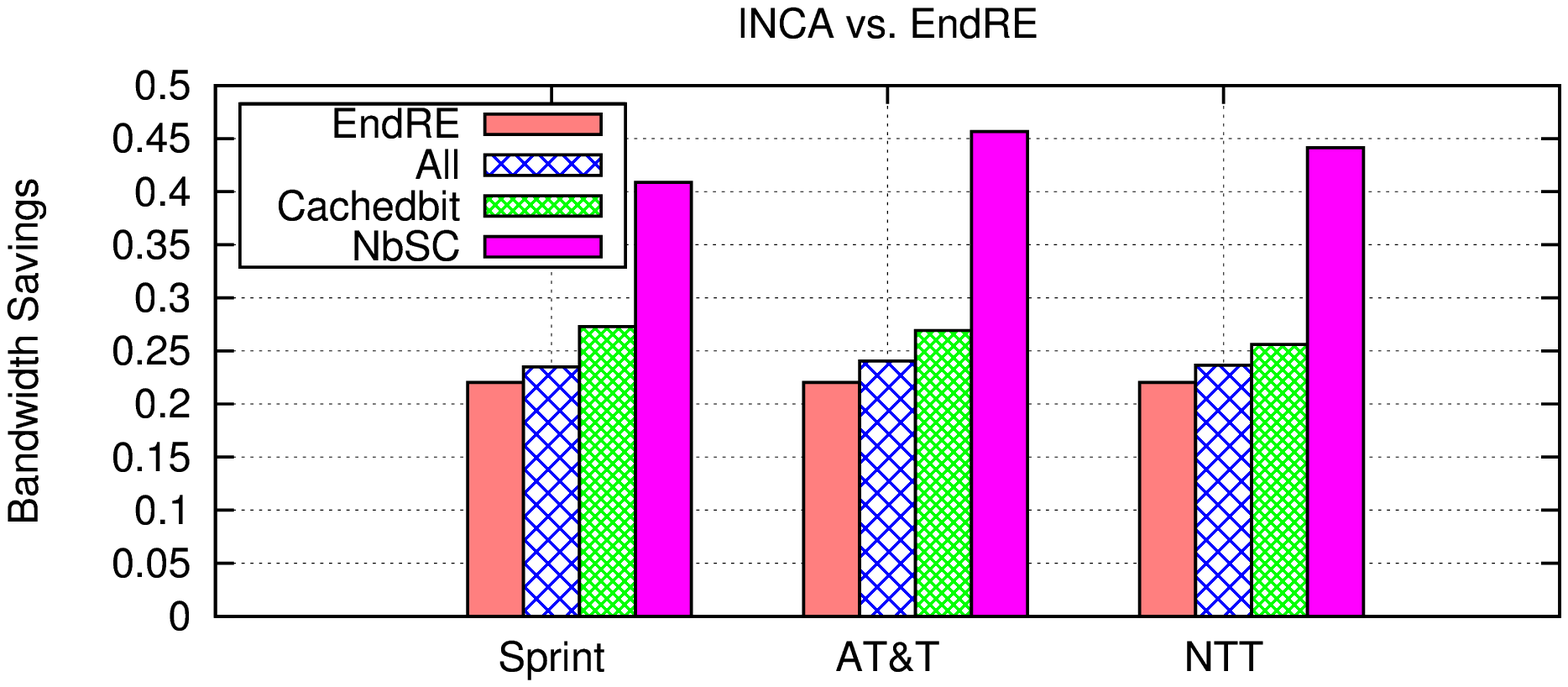}}\\
  \subfloat[256 chunks of storage]{\label{fig:inca-vs-endre-256}\includegraphics[width=7cm]{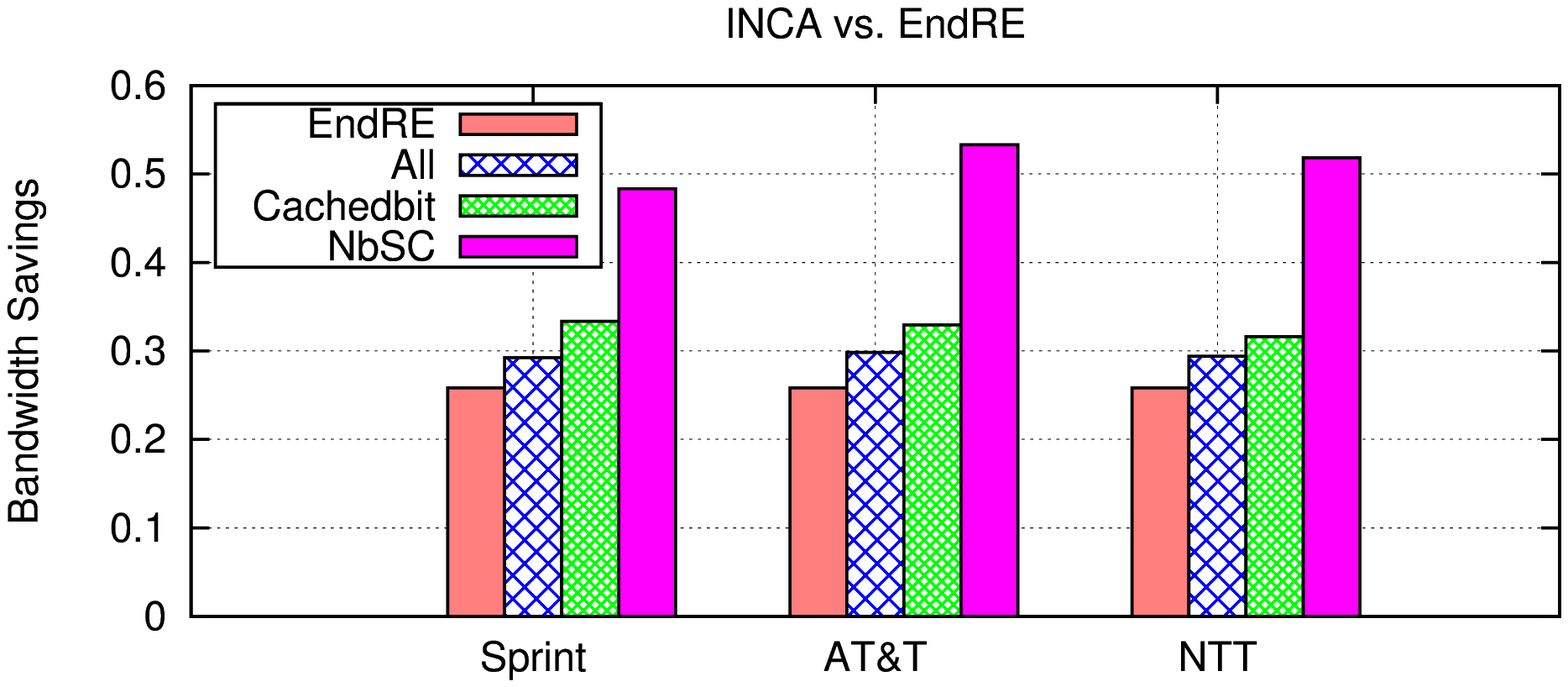}}

  \caption{INCA vs. EndRE}
  \label{fig:inca-vs-endre}
  \vskip -5mm
\end{figure}

Figure \ref{fig:inca-vs-endre} shows the bandwidth savings of both
INCA and EndRE~\cite{aggarwal2010endre} on three different
networks. We show cache sizes of 128 and 256 chunks. The bandwidth
savings of EndRE remains the same on three networks because it is
end-to-end solution. The network topology does not affect its
performance. We can clearly see that INCA is superior to EndRE. Even
the ALL strategy is slightly better than EndRE in all three
networks. PACK~\cite{zohar:pack} is another end-to-end RE solution,
but according to~\cite{zohar:pack}, its performance is about 2\% worse
than EndRE. Larger cache sizes improve INCA's performance;
figures not shown due to space limitations. Note that INCA's savings
are a combination of results shown in Figures~\ref{fig:hitrate}
and~\ref{fig:footprint}. 



Anand et al.~\cite{anand:sigmetrics} have evaluated real trace
captures and their results suggest that a middlebox-based solution
(i.e., something akin to INCA) has an advantage over end-to-end RE
solutions in saving network bandwidth. INCA does have a definite
advantage in not requiring synchronization between the server and
client and since some content can be served from CRs along the path,
we avoid having to do a round-trip to the origin of the content,
possibly speeding up the transfer.

\section{Conclusion}
\label{s:conclusion}

In this paper we have compared in-network caching with standard
redundancy elimination solutions in terms of their effectiveness at
reducing network traffic load. As an example of in-network caching, we
have presented INCA, a caching architecture which aims at capturing
the salient features of information-centric networks. We have kept the
design of INCA minimal and only consider simple solutions for the
problems of caching and routing. Our comparison on Rocketfuel
topologies shows that INCA is superior to SmartRE in the ability to
reduce external and internal network traffic, with additional
reductions of up to 65\% in internal traffic. Similar results hold for
comparisons against end-to-end RE solutions. 

\end{document}